\documentstyle[aps,prb,epsf,multicol]{revtex}

\newcommand{\bleq}{\ifpreprintsty
                   \else
                   \end{multicols}\vspace*{-3.5ex}{\tiny 
                   \noindent\begin{tabular}[t]{c|}
                   \parbox{0.493\hsize}{~} \\ \hline \end{tabular}}
                   \fi}
\newcommand{\eleq}{\ifpreprintsty
                   \else
                   {\tiny\hspace*{\fill}\begin{tabular}[t]{|c}\hline
                    \parbox{0.49\hsize}{~} \\ 
                    \end{tabular}}\vspace*{-2.5ex}\begin{multicols}{2}
                    \fi}
\newcommand{\bcols}{\ifpreprintsty\else\begin{multicols}{2}\fi}
\newcommand{\ecols}{\ifpreprintsty\else\end{multicols}\fi}

\newcommand{\tr}{\mbox{tr}\, }

\begin{document}
\draft

\title{Transmission through a many-channel random waveguide with absorption}

\author{P.\ W.\ Brouwer}
\address{Instituut-Lorentz, Leiden University, P.O. Box 9506, 2300 RA
Leiden, The Netherlands\\
and Lyman Laboratory of Physics, Harvard University, Cambridge, MA 02138, USA\cite{present}}
\date{\today}
\maketitle

\begin{abstract}
We compute the statistical distribution of the transmittance of a random waveguide with absorption in the limit of many propagating channels. We consider the average and fluctuations of the conductance $T = \mbox{tr}\, t^{\dagger} t$, where $t$ is the transmission matrix, the density of transmission eigenvalues $\tau$ (the eigenvalues of $t^{\dagger} t$), and the distribution of the plane-wave transmittances $T_a$ and $T_{ab}$. For weak absorption (length $L$ smaller than the exponential absorption length $\xi_a$), we compute moments of the distributions, while for strong absorption ($L \gg \xi_a$), we can find the complete distributions. Our findings explain recent experiments on the transmittance of random waveguides by Stoytchev and Genack [Phys.\ Rev.\ Lett.\ {\bf 79}, 309 (1997)].\smallskip
\pacs{PACS numbers: 42.25.Bs, 42.68.Mj, 41.20.Jb, 72.15.Rn}
\end{abstract}



\bcols

\section{introduction}

Interference between multiply scattered waves leads to strong fluctuations in the transmitted intensity through a disordered mesoscopic sample.\cite{AltshulerReview} While a theory of the primary fluctuation phenomena was originally constructed in the context of electron transport through metals or semiconductors,\cite{Altshuler,LeeStone} it was soon realized that the fluctuations are characteristic of systems involving multiple elastic scattering of any kind of waves, like sound, microwaves, and light, and that they can described within the same theoretical framework.\cite{Feng} Experiments with light or microwave radiation can be carried out with a very high accuracy, and allow for a precise verification of the theoretical predictions for the complete transmission distribution, rather than of its mean and variance only.\cite{GenackGarcia,DeBoer,Stoytchev} 

The relative importance of the fluctuations is determined by the ratio $N \ell/L$, where $N$ is the number of transverse propagating channels in the waveguide, $L$ its length, and $\ell$ is the elastic mean free path. 
To observe strong fluctuations, it is important to achieve as low values of $N\ell/L$ as possible. This is difficult for optical experiments, because scatterers are weak and the typical number of transverse channels $N$ is large.\cite{Maret} Moreover, in contrast to electronic systems, where the total flux is conserved, in optical systems the interference pattern may be affected as a result of loss or absorption. Absorption does not destroy the phase coherence of a transmitted wave; it merely rearranges the interference pattern, while the relative importance of the interference phenomena is unaffected. This is different from electronic systems, where inelastic scattering due to e.g.\ electron-electron interactions obscures the interference phenomena because of the presence of a large incoherent background signal.

An important step in the pursuit of low values of $N\ell/L$ for optical systems was recently reported by Stoytchev and Genack.\cite{Stoytchev} They achieved $N\ell/L \approx 3$ for microwave transmission through a copper tube with randomly placed polystyrene scatterers (a ``random waveguide''). Apart from the length-dependence of the variance of the transmittance, the measured transmittance distribution agrees surprisingly well with the theoretical predictions.\cite{NieuwenhuizenRossum,KoganKaveh,vanLangen} The agreement is surprising for a number of reasons. First, because the experiment is in a regime of strong absorption, the longest waveguide being approximately five times the exponential absorption length $\xi_a$, while the theory of Refs.\ \onlinecite{NieuwenhuizenRossum,KoganKaveh,vanLangen} was derived for waveguides without absorption. Second, because the variance of the transmittance depends sublinearly on $L$, which can not be explained within the existing theory for non-absorbing random waveguides. And third, because the experiment was compared to a theory for $N \ell/L \gg 1$, while $N \ell/L$ was not really large in the experiment.

Several papers have dealt with the problem of transmission through absorbing waveguides with only one propagating mode ($N=1$).\cite{Doucot,Freilikher,Zhang,Paasschens,Heinrichs} In order to analyze and explain optical or microwave experiments like that of Ref.\ \onlinecite{Stoytchev}, which are done in waveguides with many propagating channels, it is necessary that a theory of the transmission fluctuations and the localization transition in many-channel random waveguides with absorption ($N \gg 1$) be developed. It is the aim of the present paper to present such a theory.

\begin{figure}
\bigskip

\epsfxsize=0.89\hsize
\hspace{0.04\hsize}
\epsffile{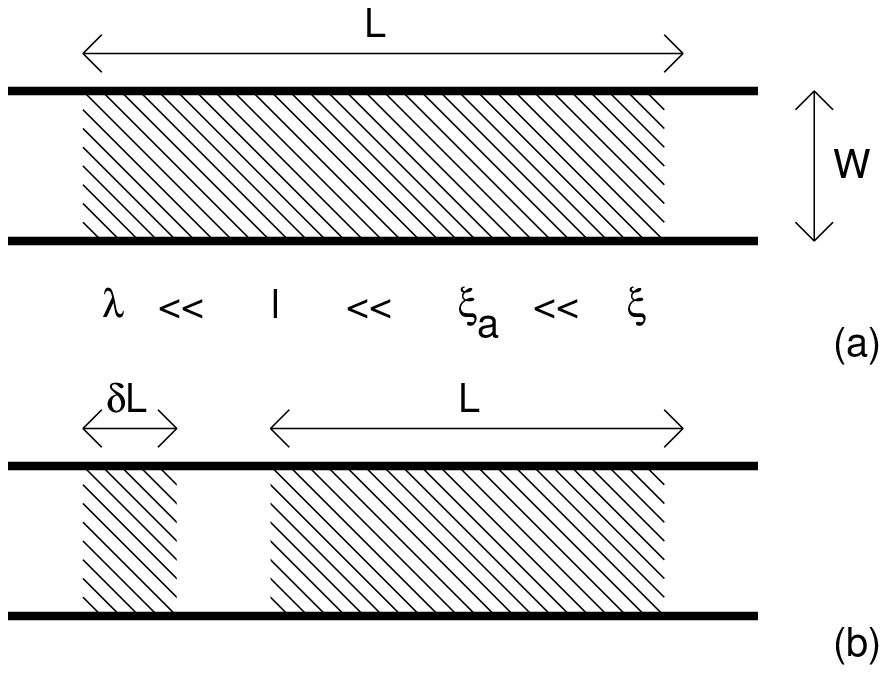}
\refstepcounter{figure} 
\label{fig:1}
\bigskip

{\small FIG.\ \ref{fig:1}. (a) The random waveguide that we consider here is quasi one-dimensional: its length $L$ is much larger than its width $W$. The other relevant length scales are the wavelength $\lambda$, the elastic mean free path $\ell$, the exponential decay length $\xi_a$, and the localization length $\xi = N \ell$, where $N \gg 1$ is the number of transverse channels in the waveguide. We assume $\lambda \ll \ell \ll \xi_a \ll \xi$. No assumption is made about the length $L$ compared to $\xi_a$ and $\xi$. (b) A thin slice of length $\delta L$ (left) is added to a random waveguide.
\par}
\end{figure}

The geometry of the random waveguide is quasi one-dimensional (width $W$ much smaller than the length $L$). The relevant length scales are the elastic mean free path $\ell$, the exponential absorption length $\xi_a$, and the localization length $\xi = N \ell$, where $N \gg 1$ is the number of transverse channels in the waveguide, see Fig.\ \ref{fig:1}a. Although a priori the localization length $\xi$ is a property of the non-absorbing system, we find that it also governs the interference effects and the localization transition in a quasi one-dimensional random waveguide with absorption. We assume that $\ell \ll \xi_a \ll \xi$, i.e.\ absorption is weak on the scale of a single scattering event, but it becomes dominant before interference effects cause waves to localize. This is appropriate for experiments on the transmission of light and microwave radiation through random waveguides.\cite{GenackGarcia,DeBoer,Stoytchev} The assumption $\xi_a \ll \xi$ is crucial for our theory of the transmittance fluctuations in the localized regime $L \gg \xi$.

Transmission through the waveguide is described by the $N \times N$ transmission matrix $t$, from which the three principal types of transmittances can be computed,
\begin{equation}
  T_{ab} = |t_{ab}|^2,\ \ T_{a} = \sum_{b=1}^{N} |t_{ab}|^2,\ \ 
  T = \sum_{a,b=1}^{N} |t_{ab}|^2.
\end{equation}
The transmittance $T$ is the equivalent of the conductance for an electronic system. It is the transmitted intensity if the sample is illuminated through a diffusor (all channels have equal incident power). The transmittances $T_{a}$ and $T_{ab}$ measure the total transmitted intensity and the intensity in channel $b$, respectively, if the sample is illuminated through channel $a$ only. (For $N \gg 1$, this corresponds to plane-wave illumination.) The transmittance $T_{ab}$ is related to the speckle pattern, the configuration of randomly positioned dark and bright spots observed if a disordered sample is illuminated by a laser beam.
The transmission matrix $t$ can be decomposed into unitary matrices $u$ and $v$ and a matrix of transmission eigenvalues $\tau_{\mu}$ ($\mu =1,\ldots,N$), the eigenvalues of $t^{\dagger} t$,
\begin{equation} \label{eq:polar}
  t = u\, \mbox{diag}\,(\tau_1^{1/2},\ldots,\tau_N^{1/2})\, v, \ \ 0 \le \tau_{\mu} \le 1.
\end{equation}
Time-reversal symmetry implies that $v = u^{\rm T}$.
For comparison with the electronic case, we also address the case of broken time-reversal symmetry. The matrices $u$ and $v$ are uniformly distributed in the unitary group,\cite{Misirpashaev} as in the absence of absorption.\cite{MelloAkkermansShapiro} To find the transmittance distribution, it remains to find the statistical distribution of the transmission eigenvalues $\tau_{\mu}$ in the limit $N \gg 1$ corresponding to thick waveguides.

This paper is organized as follows: In Sec.\ \ref{sec:2} we recall the scattering approach for the distribution of the transmittance, and its extension to absorbing systems.\cite{Paasschens,Misirpashaev} The transmittance distribution in the diffusive regime $\ell \ll L \ll \xi$ is considered in Sec.\ \ref{sec:3}. In  \ref{sec:4} we consider the crossover to the localized regime $L \gtrsim \xi$. Secs.\ \ref{sec:3} and \ref{sec:4} primarily focus on the statistical distribution of the transmittance (or conductance) $T$. The distribution of the transmittances $T_{a}$ and $T_{ab}$ is discussed in Sec.\ \ref{sec:5}. In Sec.\ \ref{sec:5b}, we discuss the relation of our work to the experiments of Stoytchev and Genack.\cite{Stoytchev} We conclude in Sec.\ \ref{sec:6}.

\section{Scattering approach} \label{sec:2}

The statistical distribution of the transmission matrix is obtained using a scattering approach similar the Fokker-Planck approach to the distribution of transmission eigenvalues in a disordered waveguide without absorption.\cite{Dorokhov,MPK,MelloStone,BeenakkerReview} The Fokker-Planck approach was also applied to the reflection eigenvalues of a disordered absorbing waveguide.\cite{Doucot,Vasilev,PradhanKumar,BPB,BruceChalker} Technical difficulties\cite{Misirpashaev} prevented a further generalization to the transmission eigenvalues of an absorbing waveguide beyond the case $N=1$.\cite{Doucot,Freilikher,Zhang,Paasschens,Heinrichs} Starting from the random-matrix model of Ref.\ \onlinecite{Misirpashaev}, we take a slightly different approach, which is explained below.

Consider a disordered waveguide with $N$ propagating channels and length $L$, see Fig.\ \ref{fig:1}. Its transmission and reflection properties are described by the scattering matrix $S$. The $2 N \times 2 N$ matrix $S$ has the standard decomposition into $N \times N$ reflection and transmission matrices,
\begin{equation}
  S = \left( \begin{array}{cc} r & t' \\ t & r' \end{array} \right).
\end{equation}
We now add a thin slice of width $\delta L$ to the waveguide and calculate the change of the transmission matrix $t$ and the reflection matrix $r$. The slice has scattering matrix $S_1$, which is parameterized as\cite{Misirpashaev}
\begin{equation} \label{eq:param}
  S_1 = \left( \begin{array}{ll} r_1 & t'_1 \\ t_1 & r'_1 \end{array} \right) =
  \left( \begin{array}{rr} v' \sqrt{\rho} v & v' \sqrt{\tau} u' \\
  u \sqrt{\tau} v & - u \sqrt{\rho} u' \end{array} \right).
\end{equation}
Here $u$, $u'$, $v$, and $v'$ are $N \times N$ unitary matrices and $\tau = \mbox{diag}\,(\tau_1, \ldots, \tau_N)$ and $\rho = \mbox{diag}\,(\rho_1, \ldots, \rho_N)$ are diagonal matrices containing the reflection and transmission eigenvalues of the thin slice. In the presence of time-reversal symmetry ($\beta=1$), one has $u' = u^{\rm T}$ and $v' = v^{\rm T}$. A statistical ensemble of disordered waveguides is obtained by considering waveguides with different configurations of the scatterers. Following Ref.\ \onlinecite{Misirpashaev}, we assume that $u$, $u'$, $v$, and $v'$ are uniformly distributed in the unitary group, and that the first moments of the diagonal matrices $\rho$ and $\tau$ are 
\begin{mathletters} \label{eq:TRslice}
\begin{eqnarray}
  N^{-1} \langle \mbox{tr}\, \rho \rangle &=& \delta L/\ell,\\
  N^{-1} \langle \mbox{tr}\, \tau \rangle &=& 1 - (\delta L/\ell + \delta L/\ell_a),
\end{eqnarray}
\end{mathletters}%
where $\ell$ is the elastic mean free path and $\ell_a$ is the ballistic absorption length. The ballistic absorption length $\ell_a$ is related to the exponential decay length $\xi_a$ as $\ell_a = 2 \xi_a^2/\ell$.

Upon addition of the thin slice at the left end of the disordered waveguide, its transmission matrix $t$ and reflection matrix $r$ are changed according to
\begin{mathletters}\label{eq:trchange}
\begin{eqnarray}
  t &\to& t (1 - r_1' r)^{-1} t_1,\\
  r &\to& r_1 + t_1' r (1 - r_1' r)^{-1} t_1.  
\end{eqnarray}
\end{mathletters}%
The new transmission and reflection matrices $t$ and $r$ do not depend on $t'$ and $r'$. Since we know the statistical distribution of the matrices $t_1$, $t_1'$, $r_1$, and $r_1'$ of the thin slice, and of the matrices $t$ and $r$ of the waveguide at length $L$, we thus can find the statistical distribution of the transmission and reflection matrices $t$ and $r$ of the waveguide at length $L + \delta L$. In this way, one obtains a Fokker-Planck equation for the distribution of $r$ and $t$.

In non-absorbing random waveguides ($\ell_a \to \infty$) this Fokker-Planck equation depends on the transmission eigenvalues $\tau_1,\ldots,\tau_N$ only. It is known as the Dorokhov-Mello-Pereyra-Kumar (DMPK) equation,\cite{Dorokhov,MPK} and it is one of the major tools for the study of quantum transport.\cite{BeenakkerReview} The DMPK equation has been generalized to the reflection eigenvalues of random waveguides with absorption.\cite{PradhanKumar,BPB,BruceChalker} For transmission through absorbing waveguides with many propagating channels ($N \gg 1$), however, the Fokker-Planck approach proves useless:\cite{Misirpashaev} The transmission eigenvalues do not decouple from the eigenvectors of $t^{\dagger} t$ and $r^{\dagger} r$, so that the number of variables is of order $N^2$, rather than $N$. In this work, we take a different approach: we use Eqs.\ (\ref{eq:TRslice}) and (\ref{eq:trchange}) to derive a set of partial differential equations for the ensemble-averages of traces of (products of) $r$ and $t$, without direct reference to the transmission eigenvalues $\tau_{\mu}$. [We need to include the reflection matrix $r$, because the $L$-evolution of $t$ depends on $r$, cf.\ Eq.\ (\ref{eq:trchange}).] A similar set of evolution equations for traces of the form $\mbox{tr}\, (t^{\dagger} t)^{n}$ for the case of non-absorbing random waveguides has been derived from the DMPK equation in Refs.\ \onlinecite{MelloStone} and from a microscopic theory in Ref.\ \onlinecite{Tartakowski}. In the next two sections we present a detailed discussion of these evolution equations and their solution in the limit of random waveguides with many channels ($N \gg 1$).

\section{Diffusive regime} \label{sec:3}

In this section we consider evolution equations for traces of products of the reflection matrix $r$ and the transmission matrix $t$, and obtain a solution as an expansion in $1/N$. We keep the ratio $L/\ell$ fixed as we expand in $1/N$. Such a expansion is valid in the diffusive regime $\ell \ll L \ll \xi$, where $\xi=N \ell$ is the localization length of the system in the absence of disorder. We explain the method by the computation of the averages $\langle T \rangle = \langle \mbox{tr}\, t^{\dagger} t \rangle$ and $\langle \mbox{tr}\, r^{\dagger} r \rangle$ to leading order in $N$, and then discuss the more general traces $\mbox{tr} (t^{\dagger} t)^{n}$, the density of transmission eigenvalues $\tau$, the weak-localization correction to the average transmittance, and the transmittance fluctuations.

\subsection{Average of the transmittance $T$}

The simplest evolution equations are those for the average transmittance $\langle T \rangle = \langle \mbox{tr}\, t^{\dagger} t \rangle$ and the average reflectance $\langle R \rangle = \langle \mbox{tr}\, r^{\dagger} r \rangle$. Combination of Eq.\ (\ref{eq:TRslice}) and (\ref{eq:trchange}) yields
\begin{mathletters} \label{eq:trevol1}
\begin{eqnarray}
  {\partial_L \langle \mbox{tr}\, t^{\dagger} t \rangle} &=&
  -\left( \ell^{-1}_{\vphantom{a}} + \ell^{-1}_{{a}} \right) \langle \mbox{tr}\, t^{\dagger} t \rangle
   + c_{\beta} \ell^{-1} \langle \mbox{tr}\, t^{\dagger} t \, \mbox{tr}\, r^{\dagger} r \rangle 
  \nonumber \\ && \mbox{} 
  + \delta_{\beta,1} c_{\beta} \ell^{-1} \langle \mbox{tr}\, t^{\dagger} t r^{\dagger} r \rangle, \label{eq:trttevol} \\
  {\partial_L \langle \mbox{tr}\, r^{\dagger} r \rangle} &=&
  -2 \left( \ell^{-1}_{\vphantom{a}} + \ell^{-1}_{{a}} \right) \langle \mbox{tr}\, r^{\dagger} r \rangle
  + c_{\beta} \ell^{-1} \langle (\mbox{tr}\, r^{\dagger} r)^2 \rangle
  \nonumber \\ && \mbox{} 
  + N 
  + \delta_{\beta,1} c_{\beta}  \ell^{-1} \langle \mbox{tr}\, (r^{\dagger} r)^2 \rangle, \label{eq:trrrevol}
\end{eqnarray}
\end{mathletters}%
where $c_{\beta} = \beta/(\beta N + 2 - \beta)$ and $\beta=1$ ($2$) in the presence (absence) of time-reversal symmetry. Although Eq.\ (\ref{eq:trevol1}) does not form a closed set of equations from which the averages $\langle \mbox{tr}\, t^{\dagger} t \rangle$ and $\langle \mbox{tr}\, r^{\dagger} r \rangle$ can be computed directly, it can be used to compute the transmittance distribution in the limit of many channels $(N \gg 1)$. The reason is that the terms that couple the $L$-dependence of $\langle \mbox{tr}\, t^{\dagger} t \rangle$ and $\langle \mbox{tr}\, r^{\dagger} r \rangle$ to averages of traces with higher powers of $t$ and $r$ are small by a factor of order $N$, so that their effect can be taken into account perturbatively. For non-absorbing waveguides, such large-$N$ expansions have been studied in Ref.\ \onlinecite{MelloStone}; for the reflection properties of absorbing waveguides, a large-$N$ solution according to these lines was given in Ref.\ \onlinecite{MisirpashaevBeenakker}. 

To find the leading large-$N$ behavior of $\langle \mbox{tr}\, t^{\dagger} t \rangle$ and $\langle \mbox{tr}\, r^{\dagger} r \rangle$, we retain only terms that are of order $N$ in the differential equation (\ref{eq:trevol1}). A trace is counted as a factor $N$. To leading order in $N$, the average of a product of traces equals the product of the averages,\cite{BeenakkerReview} while corrections are of relative order $N^{-2}$. Further, we may neglect $\langle \mbox{tr}\, t^{\dagger} t r^{\dagger} r \rangle$ and $\langle \mbox{tr}\, (r^{\dagger} r)^2 \rangle$ with respect to $\langle \mbox{tr}\, t^{\dagger} t\rangle \langle \mbox{tr}\, r^{\dagger} r \rangle$ and $\langle \mbox{tr}\, r^{\dagger} r \rangle^2$. Then Eq.\ (\ref{eq:trevol1}) simplifies to
\begin{mathletters} \label{eq:trevolsimp}
\begin{eqnarray}
  \ell\, {\partial_L \langle \mbox{tr}\, t^{\dagger} t \rangle }  &=&
  -\left( 1 + \gamma - N^{-1} \langle \mbox{tr}\, r^{\dagger} r 
    \rangle \right) \langle \mbox{tr}\, t^{\dagger} t \rangle 
  \nonumber \\ && \mbox{} + {\cal O}(1), \\
  \ell\, {\partial_L \langle \mbox{tr}\, r^{\dagger} r \rangle} &=&
  -2 \left( 1 + \gamma \right) \langle \mbox{tr}\, r^{\dagger} r \rangle
  + N + N^{-1} \langle \mbox{tr}\, r^{\dagger} r \rangle^2 
  \nonumber \\ && \mbox{} + {\cal O}(1),
\end{eqnarray}
\end{mathletters}%
where $\gamma = \ell/\ell_a$. These equations are the same as those obtained from a diffusion-equation approach, neglecting the wavelike nature of the radiation. In the next subsections, interference corrections will be taken into account by addition of the terms that we discarded as we simplified Eq.\ (\ref{eq:trevol1}) to Eq.\ (\ref{eq:trevolsimp}). The solution of Eq.\ (\ref{eq:trevolsimp}) with initial conditions $\langle \mbox{tr}\, t^{\dagger} t \rangle = N$, $\langle \mbox{tr}\, r^{\dagger} r \rangle = 0$ at $L=0$ reads for $L, \ell_a \gg \ell$
\begin{mathletters} \label{eq:avg}
\begin{eqnarray}
  \langle T \rangle  = \langle \mbox{tr}\, t^{\dagger} t \rangle &=& {\xi \over \xi_a \sinh s},
  \\
  \langle R \rangle = \langle \mbox{tr}\, r^{\dagger} r \rangle &=& N - {\xi \over \xi_a} \coth s,
\end{eqnarray}
\end{mathletters}
where $\xi_a = [\ell \ell_a/2]^{1/2}$ and $s = L/\xi_a$. The length scale $\xi_a$ is the classical exponential decay length for an absorbing random waveguide. In the weak absorption regime $L \ll \xi_a$, Eq.\ (\ref{eq:avg}) simplifies to Ohm's law $\langle T \rangle = N - \langle R \rangle = \xi/L$, while in the strong absorption regime $L \gg \xi_a$, the reflectance $R$ saturates at the value $N - \xi/\xi_a$, while the transmittance $T$ decays exponentially with decay length $\xi_a$,
\begin{equation} \label{eq:tavgN}
  \langle T \rangle = {2 \xi \over \xi_a} e^{-L / \xi_a},\ \ L \gg \xi_a.
\end{equation}

\subsection{Traces of the form $\mbox{tr}\, (t^{\dagger} t)^{n}$ and the density of transmission eigenvalues}

We now generalize the evolution equation (\ref{eq:trevolsimp}) to arbitrary traces of the form
\begin{equation}
  M_{x_1,\ldots,x_n} \equiv \langle \mbox{tr}\, x_1^{\dagger} x_1 \ldots x_n^{\dagger} x_n \rangle,
\end{equation}
where the symbol $x_j$ can be $t$ or $r$. These traces are important for the density of transmission eigenvalues and for the distribution of the transmittances $T_{a}$ and $T_{ab}$ in the diffusive regime, see Sec.\ \ref{sec:5}. Repeating the steps leading to Eq.\ (\ref{eq:trevolsimp}), we find \bleq
\begin{eqnarray}
  \ell\, {\partial_L} M_{x_1,\ldots,x_n} &=& 
  -(p_1+1) \left( 1 + \gamma \right) M_{x_1,\ldots,x_n}
  \nonumber \\ && \mbox{} 
  +
  \sum_{k=1}^{n} {1 \over N} M_{x_1,\ldots,x_k} M_{r,x_{k+1},\ldots,x_{n}}
  +
  \sum_{k=1}^{n} {p_1 p_k \over N} 
    M_{x_1,\ldots,x_{k-1}} M_{x_{k+1},\ldots,x_{n}}
  \nonumber \\ && \mbox{} -
  \sum_{k=2}^{n} {p_1+p_n \over N}
    M_{x_1,\ldots,x_{k-1}} M_{x_{k},\ldots,x_{n}}
  \nonumber \\ && \mbox{}
   + \mbox{cyclic permutations} + {\cal O}(1), \label{eq:evol} \\
  M_{x_1,\ldots,x_n}(0) &=& N \prod_{j=1}^{n} (1-p_j), \nonumber
\end{eqnarray}
where $p_j = 1$ ($0$) if $x_j$ is $r$ ($t$). The solution for $L, \ell_a \gg \ell$ and $n=2$ reads
\begin{mathletters} \label{eq:MomAbs}
\begin{eqnarray}
  \langle \mbox{tr}\, (t^{\dagger} t)^2 \rangle &=&
  {\xi \over \xi_a} \left({2 s + \coth s \over 4 \sinh^2 s} - {s \over 4 \sinh^4 s} \right), \\
  \langle \mbox{tr}\, t^{\dagger} t r^{\dagger} r \rangle &=&  
  {\xi \over \xi_a} \left({1 \over 4 \sinh s} + {s \coth s - 1 \over 4 \sinh^3 s} \right), \\
  \langle \mbox{tr}\, (r^{\dagger} r)^2 \rangle &=& N
  - {\xi \over \xi_a} \left( {3 \coth s \over 2} + {\coth s \over 4 \sinh^2 s} - {s\over 4 \sinh^4 s} \right), 
\end{eqnarray}
\end{mathletters}%
\eleq\noindent
where as before $s = L/\xi_a$. For weak absorption, Eq.\ (\ref{eq:MomAbs}) agrees with results obtained from the DMPK equation,\cite{MelloStone} while for strong absorption Eq.\ (\ref{eq:MomAbs}) simplifies to 
\begin{eqnarray}
  \langle \mbox{tr}\, (t^{\dagger} t)^2 \rangle &=& 2 (L \xi/\xi_a^2) e^{-2 L/\xi_a}, \nonumber \\
  \langle \mbox{tr}\, t^{\dagger} t r^{\dagger} r \rangle &=&  (\xi/2 \xi_a) e^{-L/\xi_a}, \label{eq:mom2} \\
  \langle \mbox{tr}\, (r^{\dagger} r)^2 \rangle &=&
  N - (3 \xi/2 \xi_a). \nonumber 
\end{eqnarray}

The averages $\langle \mbox{tr}\, (t^{\dagger} t)^{n} \rangle$ correspond to moments of the density $\rho(\tau)$ of transmission eigenvalues $\tau$. The density $\rho(\tau)$ is recovered from the moments as the imaginary part of the Green function $G(z)$
\begin{mathletters} 
\begin{eqnarray}
  \rho(\tau) &=& \pi^{-1} \mbox{tr}\, G(\tau + i0), \nonumber \\
  G(z) &=& 
  \left\langle \mbox{tr}\, (z - t^{\dagger} t)^{-1} \right\rangle 
 = \label{eq:G}
  \sum_{n=0}^{\infty} 
    \left\langle {\tr (t^{\dagger} t)^{n} \over z^{n+1}}\right\rangle.
\end{eqnarray}
\end{mathletters}%
For the calculation of $\rho(\tau)$ we thus need the moments $M_{t,\ldots,t} = N^{-1} \mbox{tr}\, (t^{\dagger} t)^{n}$ for all $n$. In principle this requires computation of all moments $M_{x_1,\ldots,x_m}$ with $m \le n$ and with $x_j$ being either $r$ or $t$. As the number of possible moments $M_{x_1,\ldots,x_m}$ proliferates exponentially fast with increasing $m$, this is not feasible. This is a fundamental difference with the set of moment equations for the case of a non-absorbing waveguide, where only moments of the form $\langle \mbox{tr}\, (t^{\dagger} t)^{m} \rangle$ need to be taken into account.

In the strong absorption regime $L \gg \xi_a$, however, a solution for $\rho(\tau)$ can be found in closed form. Inspection of the general evolution equation (\ref{eq:evol}) for $L \gg \xi_a$ shows that the leading behavior of $\mbox{tr}\, (t^{\dagger} t)^{n}$ and $\mbox{tr}\, (t^{\dagger} t)^{n} r^{\dagger} r$ has the asymptotic form
\begin{mathletters}
\begin{eqnarray}
  \mbox{tr}\, \langle (t^{\dagger} t)^{n} \rangle &=& 
    a_n \xi L^{n-1} \xi_a^{-n} e^{-n L/\xi_a} + \ldots, \\
  \langle \mbox{tr}\, (t^{\dagger} t)^{n} r^{\dagger} r \rangle &=& 
    b_n \xi L^{n-1} \xi_a^{-n} e^{-n L/\xi_a} + \ldots,
\end{eqnarray}
\end{mathletters}%
where $a_{n}$ and $b_{n}$ are numerical coefficients. The dots indicate terms that are smaller by a large factor $\xi_a/L$ or $\exp(-L/\xi_a)$. The evolution equation (\ref{eq:evol}) provides a recursion relation between the coefficients $a_n$ and $b_n$,
\begin{equation}
  a_{n} = {n \over n-1} \sum_{m=1}^{n-1} b_{n-m} a_{m},\ \ 
  b_{n} = {1 \over 4} a_n . \label{eq:recurs}
\end{equation}
{}From Eq.\ (\ref{eq:tavgN}) we find $a_1 = 2$. The generating function $F(z) = \sum_{n=1}^{\infty} a_n z^{n}$ of the coefficients $a_n$ is the so-called ``product-log'' function (the principal value of the functional inverse of $x \to x e^{x}$),
\begin{equation} \label{eq:Fgener}
  F(z) = \sum_{n=1}^{\infty} a_n z^{n} = - {2}\, \mbox{Plog}\,\left({-z} \right).
\end{equation}
The product-log function is real only if its argument is larger than $-1/e$. Using Eq.\ (\ref{eq:G}), we find the Green function $G(z)$, 
\begin{eqnarray}
  G(z) &=& 
  {N \over z} - {2 \xi \over z L} \mbox{Plog}\, \left(- {L e^{-L/\xi_a} \over \xi_a z} \right),
\end{eqnarray}
and hence the density of transmission eigenvalues $\rho(\tau)$. 
\begin{figure}
\epsfxsize=0.9\hsize
\epsffile{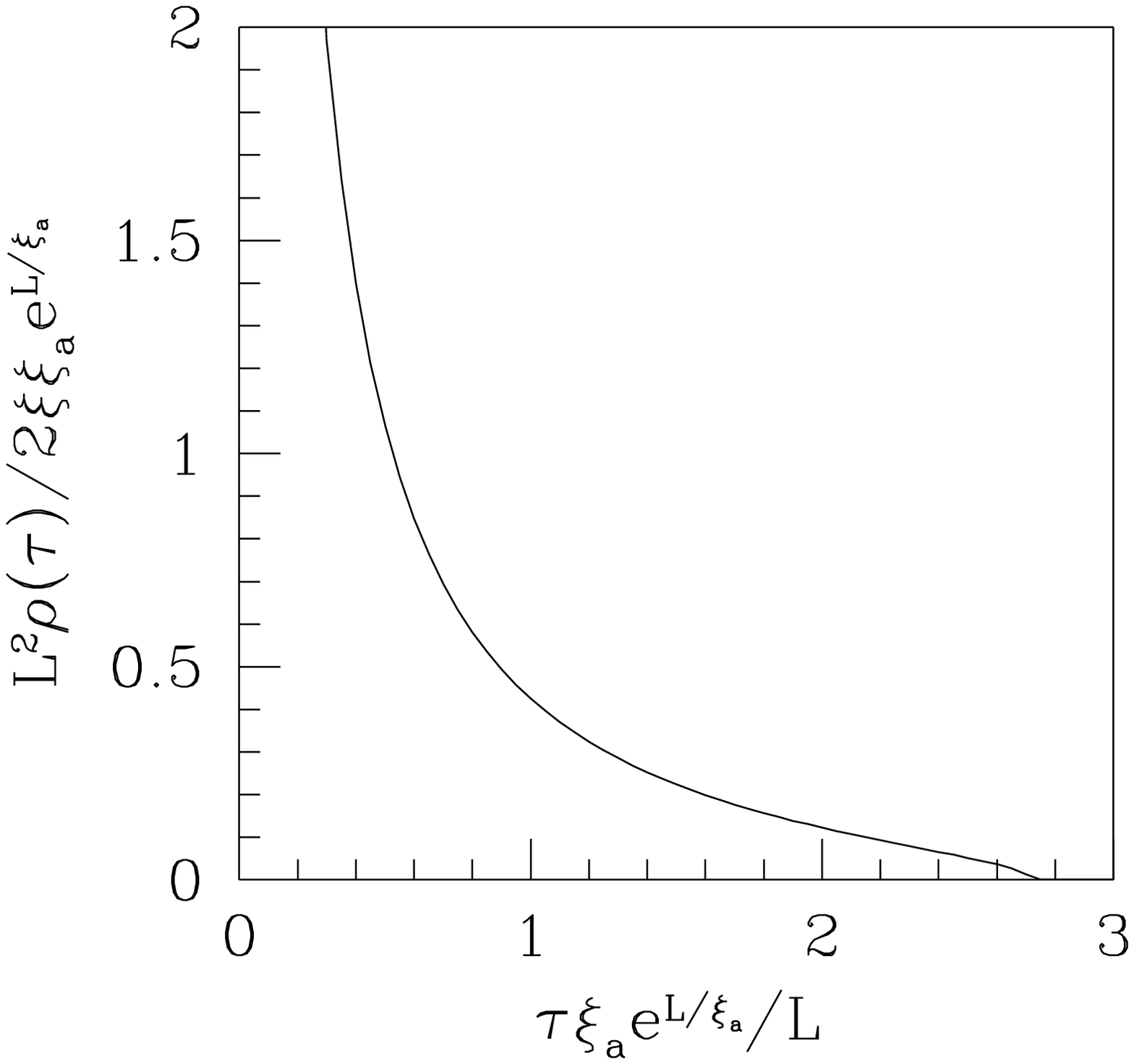}
\refstepcounter{figure}
\label{fig:2}
{\small FIG. \ref{fig:2}. Density of transmission eigenvalues of a strongly absorbing random waveguide in the diffusive regime $\xi_a \ll L \ll \xi$.}
\end{figure}

The density of transmission eigenvalues is shown in Fig.\ \ref{fig:2}. The existence of a maximum transmission eigenvalue, $\tau_{\rm max} = e L e^{-L/\xi_a}/\xi_a \ll 1$ is quite different from the case of a non-absorbing random waveguide, where the support of $\rho(\tau)$ extends from $0$ to $1$ throughout the diffusive regime.\cite{BeenakkerReview} A common feature of absorbing and non-absorbing random waveguides is that the maximal transmission eigenvalue $\tau_{\rm max}$ is a factor $\sim L/\ell \gg 1$ larger than the average transmission eigenvalue. It is this broad support of the density $\rho(\tau)$ that is responsible for many of the qualitative similarities of interference phenomena in strongly absorbing and non-absorbing systems.

\subsection{Weak localization correction}

Weak localization is a small negative interference correction to the classical transmittance.\cite{Anderson,Gorkov} In electronic systems, the weak localization correction is suppressed by a time-reversal symmetry breaking magnetic field. In the framework of an expansion in $1/N$, the weak localization correction is the ${\cal O}(1)$ correction to an ${\cal O}(N)$ average. Here we compute the weak-localization corrections $\delta T$ and $\delta R$ to the average transmittance $\langle T \rangle = \langle \mbox{tr}\, t^{\dagger} t \rangle $ and reflectance $\langle R \rangle = \langle \mbox{tr}\, r^{\dagger} r \rangle$.

Starting point is the exact evolution equation (\ref{eq:trevol1}), where we now keep all terms up to sub-leading order in $N$,
\begin{mathletters} \label{eq:trwlevol}
\begin{eqnarray}
  \ell\, {\partial_L \langle \mbox{tr}\, t^{\dagger} t \rangle }  &=&
  -\left( 1 + \gamma - N^{-1} \langle \mbox{tr}\, r^{\dagger} r \rangle \right) \langle \mbox{tr}\, t^{\dagger} t \rangle
  \nonumber \\ && \mbox{} -\delta_{\beta,1} N^{-2} \langle \mbox{tr}\, r^{\dagger} r \rangle \langle \mbox{tr}\, t^{\dagger} t \rangle 
  \nonumber \\ && \mbox{} 
  +\delta_{\beta,1} N^{-1} \langle \mbox{tr}\, t^{\dagger} t r^{\dagger} r \rangle \\
  \ell\, {\partial_L \langle \mbox{tr}\, r^{\dagger} r \rangle} &=&
  -2 \left( 1 + \gamma \right) \langle \mbox{tr}\, r^{\dagger} r \rangle
  + 1 + N^{-1} \langle \mbox{tr}\, r^{\dagger} r \rangle^2
  \nonumber \\ && \mbox{} - \delta_{\beta,1} N^{-2} \langle \mbox{tr}\, r^{\dagger} r \rangle^2
  \nonumber \\ && \mbox{} 
  + \delta_{\beta,1} N^{-1} \langle \mbox{tr}\, (r^{\dagger} r)^2 \rangle.
\end{eqnarray}
\end{mathletters}%
We substitute the result of the previous subsection for $\langle \mbox{tr}\, t^{\dagger} t r^{\dagger} r \rangle$ and $\langle \mbox{tr}\, (r^{\dagger} r)^{2} \rangle$ and solve Eq.\ (\ref{eq:trwlevol}) for $\langle \mbox{tr}\, t^{\dagger} t \rangle$ and $\langle \mbox{tr}\, r^{\dagger} r \rangle$ up to order unity. As a result, we find the weak-localization corrections to the average reflection and transmission,
\begin{mathletters}
\begin{eqnarray}
  \delta T &=& \delta_{\beta,1} \left( {\coth s -2 s \over 4 \sinh s} - {s \over 4 \sinh^3 s} 
\right), \\
  \delta R &=& \delta_{\beta,1} \left( {1 \over 4} + { s \coth s - 1 \over 4 \sinh^2 s} 
\right).
\end{eqnarray}
\end{mathletters}%
The weak localization correction to the reflectance $R$ has been calculated previously in Ref.\ \onlinecite{MisirpashaevBeenakker}. In the weak absorption regime, we recover the well known universal value $\delta R = - \delta T = \case{1}{3} \delta_{\beta,1}$,\cite{MelloStone} while for strong absorption the weak localization correction reads
\begin{eqnarray}\label{eq:tweak}
  \delta T = -\delta_{\beta,1} {L \over \xi_a} e^{-L/\xi_a},\ \
  \delta R = {1 \over 4}\delta_{\beta,1}  \label{eq:rweak}.
\end{eqnarray}

\subsection{Mesoscopic fluctuations}

Mesoscopic fluctuations of the transmittance $T$ and the reflectance $R$ are characterized by the variances $\mbox{var}\, T$, $\mbox{var}\, R$ and the covariance $\mbox{cov}\,(R,T)$. Like the weak-localization correction, they are of order $1$ in a large-$N$ expansion.\cite{BeenakkerReview} Proceeding as in the case of the weak-localization correction, we find that $\mbox{var}\, T$, $\mbox{var}\, R$ and $\mbox{cov}\,(R,T)$ obey
\bleq
\begin{mathletters} \label{eq:var}
\begin{eqnarray}
  \ell\, \partial_L \mbox{var}\, T &=& -2\, (1 + \gamma - N^{-1} \langle \mbox{tr}\, r^{\dagger} r \rangle)\, \mbox{var}\, T 
  \nonumber \\ && \mbox{} + 2 N^{-1} \langle \mbox{tr}\, t^{\dagger} t \rangle \mbox{cov}\,(R,T) + 2 N^{-1} \langle \mbox{tr}\, (t^{\dagger} t)^2 r^{\dagger} r \rangle \nonumber \\ &&
  \mbox{} + 2 \delta_{\beta,1} N^{-1} \langle \mbox{tr}\, t^{\dagger} t r^{\dagger} t^{\rm T} t^{*} r \rangle, \\
  \ell\, \partial_L \mbox{cov}\,(R,T) &=& 
  -3\, (1 + \gamma - N^{-1} \langle \mbox{tr}\, r^{\dagger} r \rangle)\, \mbox{cov}\,(R,T)
  \nonumber \\ && \mbox{} 
  + N^{-1} \langle \mbox{tr}\, t^{\dagger} t \rangle\, \mbox{var}\, R 
  - 4 \beta^{-1} N^{-1} \langle \mbox{tr}\, t^{\dagger} t r^{\dagger} r \rangle 
  \nonumber \\ && \mbox{} 
  + 4 N^{-1} \beta^{-1} \langle \mbox{tr}\, t^{\dagger} t (r^{\dagger} r)^2 \rangle, \\
  \ell\, \partial_L \mbox{var}\, R &=& -4\, (1 + \gamma - N^{-1} \langle \mbox{tr}\, r^{\dagger} r \rangle)\, \mbox{var}\, R
  \nonumber \\ && \mbox{} + 4 \beta^{-1}\, N^{-1} \langle \mbox{tr}\, r^{\dagger} r (1 - r^{\dagger} r)^2 \rangle.
\end{eqnarray}
\end{mathletters}%
The averages of the form $\langle \mbox{tr}\, (t^{\dagger} t)^n (r^{\dagger} r)^m \rangle$, can be computed from Eq.\ (\ref{eq:evol}). In the presence of time-reversal symmetry, we also need to know $\langle \mbox{tr}\, t^{\dagger} t r^{\dagger} t^{\rm T} t^{*} r \rangle$, which satisfies (for $\beta = 1$)
\begin{eqnarray} \label{eq:trb1}
  \ell\, \partial_L \langle \mbox{tr}\, t^{\dagger} t r^{\dagger} t^{\rm T} t^{*} r \rangle &=& - 4(1 + \gamma - N^{-1} \langle \mbox{tr}\, r^{\dagger} r \rangle) \langle \mbox{tr}\, t^{\dagger} t r^{\dagger} t^{\rm T} t^{*} r \rangle \nonumber \\ && \mbox{} +
  N^{-1} \langle \mbox{tr}\, t^{\dagger} t \rangle (2 \langle \mbox{tr}\, t^{\dagger} t (r^{\dagger} r)^2 \rangle - 4 \langle \mbox{tr}\, t^{\dagger} t r^{\dagger} r \rangle + \langle \mbox{tr}\, t^{\dagger} t \rangle) 
  \nonumber \\ && \mbox{} 
  + 3 N^{-1} \langle \mbox{tr}\, t^{\dagger} t r^{\dagger} r \rangle^2.
\end{eqnarray}
The solution of Eqs.\ (\ref{eq:var}) and (\ref{eq:trb1}) reads for $L, \ell_a \gg \ell$
\begin{mathletters} \label{eq:VarTcomplete}
\begin{eqnarray}
  \mbox{var}\, T &=& {2 s^2 - 9 s \coth s + 12 \over 16 \sinh^2 s} - {3 s^2 \over 16 \sinh^4 s} \nonumber \\ && \mbox{} + {1 \over \beta} \left[ {8 s \coth s - 11 \over 16 \sinh^2 s} + {6 s^2 - 3 s \coth s - 3 \over 16 \sinh^4 s} + {3 s^2 \over 8 \sinh^6 s} \right], \\
  \mbox{cov}\, (R,T) &=& {1 \over \beta} \left[{3 \coth s \over 16 \sinh s} - {2 s^2 \coth s - 5 s - 3 \coth s \over 16 \sinh^3 s} - {3 s^2 \coth s - 3 s \over 16 \sinh^5 s} \right], \\
  \mbox{var}\, R &=& {1 \over \beta} \left[ {1 \over 8} - {1 \over 16 \sinh^2 s} + {4 s^2 - 3 s \coth s - 3 \over 16 \sinh^4 s} + {3 s^2 \over 8 \sinh^6 s} \right].
\end{eqnarray}
\end{mathletters}%
\eleq\noindent
\begin{figure}
\epsfxsize=0.9\hsize
\epsffile{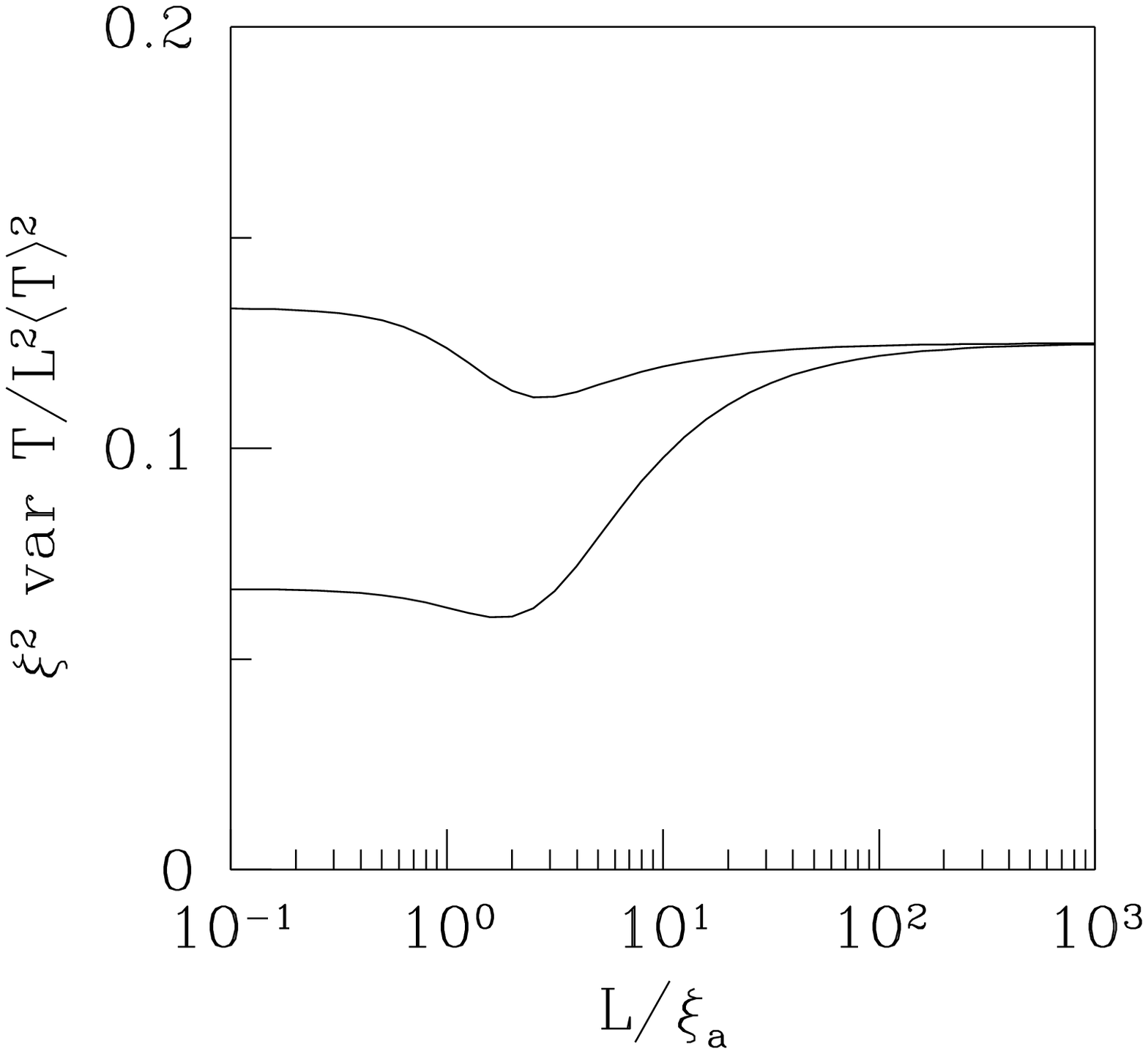}
\refstepcounter{figure}
\label{fig:VarT}
{\small FIG. \ref{fig:VarT}. Fluctuations of the transmittance $T$ in the diffusive regime $L \ll \xi$.}
\end{figure}
The $L$-dependence of the transmittance fluctuations, normalized to the average, is shown in Fig.\ \ref{fig:VarT}. For weak absorption, $L \ll \xi_a$, we find the well known universal conductance fluctuations $\mbox{var}\, T = -\mbox{cov}\,(R,T) = \mbox{var}\, R = 2/15 \beta$, while for strong absorption Eq.\ (\ref{eq:VarTcomplete}) simplifies to
\begin{eqnarray}
  \mbox{var}\, T &=& {L^2 \over 2\xi_a^2} e^{-2 L/\xi_a}, 
  \nonumber \\
  \mbox{cov}(R,T) &=&
  -{3 \over 4 \beta} e^{-L/\xi_a}, \label{eq:trcov} \label{eq:trvar} \label{eq:ttvar} \label{eq:rrvar}\\
  \mbox{var}\, R &=& {1 \over 8 \beta} \nonumber
\end{eqnarray}
The variance of $R$ was obtained earlier in Ref.\ \onlinecite{MisirpashaevBeenakker}. 
Note that the relative size of the transmittance fluctuations is equal in strongly absorbing and weakly absorbing systems, $\mbox{var}\, T \propto (L/\xi)^2 \langle T \rangle^2$.

The dependence of the fluctuations on the presence or absence of time-reversal symmetry is different from what one expects from the non-absorbing case and deserves some discussion. In non-absorbing disordered systems, the variance of the transmittance is decreased by a factor $2$ if time-reversal symmetry is broken, $\mbox{var}\, T = 2/15 \beta$. This universal $1/\beta$ dependence is well understood in terms of diagrammatic perturbation theory\cite{Altshuler,LeeStone} or random-matrix theory.\cite{DysonMehta} For a strongly absorbing system, in contrast, the size of the transmittance fluctuations does not depend on the presence or absence of time-reversal symmetry, see Eq.\ (\ref{eq:ttvar}). This is remarkable, since the average transmittance in strongly absorbing waveguides exhibits the usual $\beta$-dependent weak localization correction that is suppressed if time-reversal symmetry is broken.

\begin{figure}
\bigskip

\epsfxsize=0.89\hsize
\hspace{0.00\hsize}
\epsffile{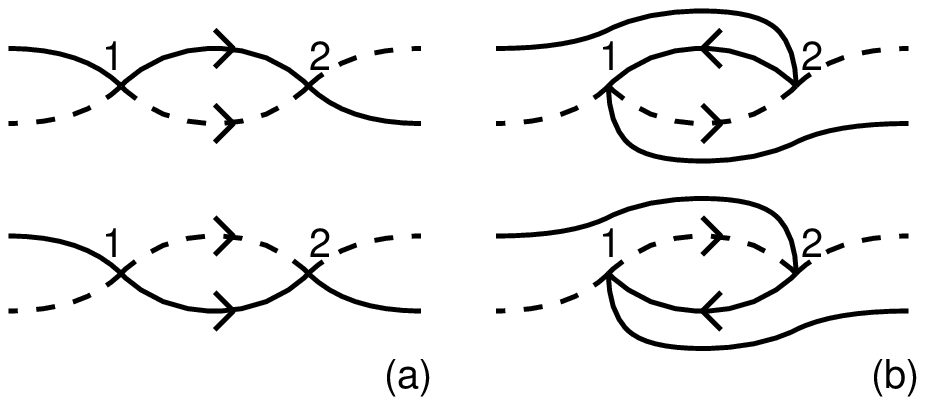}
\refstepcounter{figure} 
\label{fig:Feynman}
\bigskip

{\small FIG.\ \ref{fig:Feynman}. Feynman paths for the diffuson (a) and cooperon (b) contributions of the variance of the transmittance $T$. There is only a cooperon contribution to $\mbox{var}\, T$ if the vertices $1$ and $2$ are within a distance $\xi_a$ of each other. For diffusons, the vertices can be arbitrarily far apart. Hence for $L \gg \xi_a$, only the diffuson contribution to $\mbox{var}\, T$ survives.\par}
\end{figure}

\begin{figure}
\bigskip

\epsfxsize=0.89\hsize
\hspace{0.00\hsize}
\epsffile{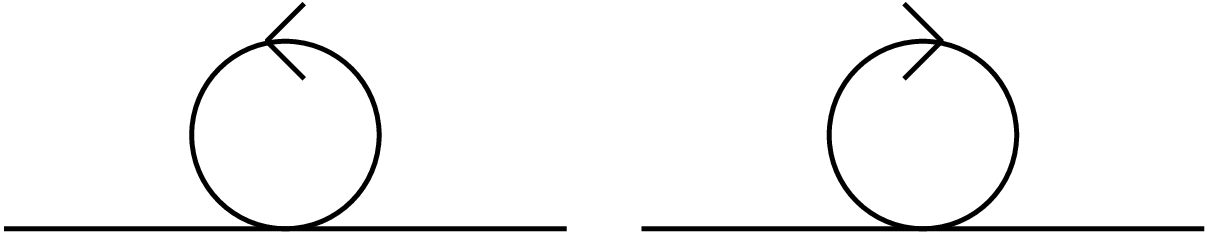}
\refstepcounter{figure} 
\label{fig:Cooperon}
\bigskip

{\small FIG.\ \ref{fig:Cooperon}. Two interfering Feynman paths for the weak-localization correction to the average transmittance $T$. As the typical length of the closed loop is of order $\ell \ll \ell_a$, the weak-localization correction is not suppressed by absorption. \par}
\end{figure}

To see why the $\mbox{var}\, T$ does not depend on the presence or absence of time-reversal symmetry in absorbing systems, we consider the relevant pairs of the Feynman paths for the diffuson and cooperon contributions to $\mbox{var}\, T$. They are shown in Fig.\ \ref{fig:Feynman}. In a waveguide without absorption, both the diffuson and cooperon paths have equal weight. The cooperon is suppressed if time-reversal symmetry is broken, thus explaining the factor two reduction of the fluctuations. 
In the presence of absorption, the diffuson and cooperon contributions no longer have equal weights, as paths for the cooperon contribution typically have a larger length. Therefore, unless the vertices $1$ and $2$ are within a distance $\xi_a$, paths of the cooperon type are suppressed. Since the diffuson paths of Fig.\ \ref{fig:Feynman}a can have vertices arbitrarily far apart, the cooperon contribution to $\mbox{var}\, T$ is a factor $\xi_a/L$ smaller than the diffuson one. This explains why the breaking of time-reversal symmetry has no effect on the transmittance fluctuations. The situation is different for the weak-localization correction to the average transmittance. The appropriate Feynman paths, which are shown in Fig.\ \ref{fig:Cooperon}, contain a closed loop with a typical length of order $\ell \ll \ell_a$. As a result the weak-localization correction to the average transmittance $\langle T \rangle$ is not suppressed by absorption. 

\subsection{Breakdown of the large-$N$ expansion}

Throughout this section we have assumed that interference corrections are small, so that we can treat them perturbatively. This assumption is bound to break down as the length $L$ of the waveguide increases, since the relative size of the interference corrections to transmission properties increases with $L$. For the transmittance $T$, both the fluctuations and the weak-localization correction become comparable to the average when the length of the waveguide approaches the localization length $\xi = N \ell$, cf.\ Eqs.\ (\ref{eq:tweak}) and (\ref{eq:ttvar}). One verifies that higher order corrections in a $1/N$ expansion become comparable to $\langle T \rangle$ as well as $L \to \xi$. This breakdown of perturbation theory has the same origin as the corresponding breakdown of perturbation theory in systems without absorption: it signals the onset of localization. A theory of the transmittance distribution in the regime $L \gtrsim \xi$ where the perturbation theory of this section is not valid, is presented in the next section. 

\section{Crossover to localized regime} \label{sec:4}

The large-$N$ perturbation theory of the previous section breaks down as the length $L$ approaches the localization length $\xi$. For non-absorbing random waveguides, a theory of the crossover from the diffusive regime into the localized regime requires a true technical tour de force,\cite{Zirnbauer,BeenakkerRejaeiExact,Rejaei} because of the intrinsically non-perturbative nature of the crossover. Although the situation for random waveguides with absorption looks similar --- it has the same divergence of perturbation theory as $L \sim \xi$ ---, it is not. The reason is the existence of the small parameter $\xi_a/\xi$, or $1/\gamma N^2$. (Note that the ratio $\xi_a/\xi$ depends on the absorption properties of the waveguide and its width, but not on its length.) In this section, we use the smallness of $\xi_a/\xi$ to compute the distribution of the transmittance for lengths $L$ comparable to, or greater than the localization length $\xi$.

A crucial distinction between the case of absorbing and non-absorbing waveguides is the saturation of the reflectance distribution for absorbing waveguides for $L \gtrsim \xi_a$. Although the relative size of the transmittance fluctuations keeps growing as $L$ exceeds the absorption length $\xi_a$ --- eventually causing the breakdown of perturbation theory ---, the correlations between reflection and transmission properties saturate. In particular, we find that
\begin{itemize}
\item  Correlations between reflection and transmission are smaller than the product of the averages by a factor $\sim \xi_a/N \xi$, irrespective of the length $L$. Hence they can be neglected if $\xi_a \ll \xi$. (For non-absorbing waveguides, correlations are smaller by a factor $L/N \xi$, and can not be ignored for $L \sim \xi$.)
\item  Traces $\mbox{tr}\, (t^{\dagger} t)^{n} r^{\dagger} (t^{\rm T} t^{*})^{m} r$ are a factor $\sim \xi_a/\xi$ smaller than $\mbox{tr}\, (t^{\dagger} t)^{n+m}$ or $\mbox{tr}\, (t^{\dagger} t)^{n} \mbox{tr}\, (t^{\dagger} t)^{m}$.
\end{itemize}
A derivation of these properties from the appropriate evolution equations is straightforward. We find it more instructive, however, to give an argument in terms of Feynman paths. 
\begin{figure}
\bigskip

\epsfxsize=0.89\hsize
\hspace{0.00\hsize}
\epsffile{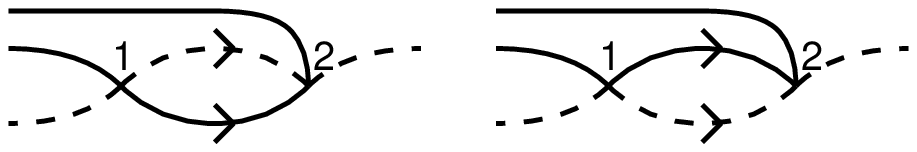}
\refstepcounter{figure} 
\label{fig:RTcorr}\bigskip

{\small FIG.\ \ref{fig:RTcorr}. Feynman paths contributing to correlations between the reflectance $R$ (solid path) and the transmittance $T$ (dashed path). The solid path, which contributes to the reflectance $R$, does not penetrate the sample further than the decay length $\xi_a$. Therefore, correlations between $R$ and $T$ saturate at $L \gtrsim \xi_a$.\par}
\end{figure}

Feynman paths giving rise to correlations between $R$ and $T$ are shown in Fig.\ \ref{fig:RTcorr}. The solid path, which contributes to $R$,  does not penetrate the sample more than a decay length $\xi_a$. As a consequence, the relative size of the correlator $\mbox{cov}(R,T)$ saturates at $L \sim \xi_a$. At this length scale, perturbation theory is still valid, and we find $\mbox{cov}(R,T) \sim (\xi_a/N \xi) \langle R \rangle \langle T \rangle \ll \langle R \rangle \langle T \rangle$ irrespective of length. The same argument applies to arbitrary correlators involving $\mbox{tr}\, (r^{\dagger} r)^{n}$, since their relative sizes saturate at $L \sim \xi_a$ as well, and to the traces of the form $\mbox{tr}\, (t^{\dagger} t)^{n} r^{\dagger} (t^{\rm T} t^{*})^{m} r$.

The formal framework in which the crossover to the localized regime is described, is the so-called ``thick-wire limit'', in which the limit $N \to \infty$ is taken, keeping $\sigma = L/\xi$ fixed. In the previous section, the ratio $L/\ell$ was kept fixed, rather than $L/\xi$. The thick-wire limit is used in the field-theoretic description of localization of in disordered waveguides.\cite{Zirnbauer,Efetov1983} Note that the thick-wire limit is unable to address length scales $L \lesssim \xi_a$ corresponding to the onset of absorption, since it corresponds to $\sigma = \xi_a/\xi \to 0$ if $N \to \infty$. In order to remove the classical exponential decay of the transmittance due to absorption, we consider traces of the form
\begin{equation} \label{eq:Tndef}
  F_n = \left(2 \xi/\xi_a \right)^{-n} e^{n L/\xi_a}\, \mbox{tr}\, (t^{\dagger} t)^{n}.
\end{equation}

The differential equations for the $\sigma$-dependence of the averages of the traces $F_n$ are derived from the exact evolution equations for the averages of (products of) traces $\mbox{tr}\, (t^{\dagger} t)^{n} (r^{\dagger} r)^{m}$, which are constructed as discussed in Sec.\ \ref{sec:2}. As an example, we consider the evolution equation for $\langle F_1 \rangle$ in the thick-wire limit. Hereto we rewrite the evolution equation (\ref{eq:trevol1}) for $\langle \mbox{tr}\, t^{\dagger} t \rangle$ in terms of $\sigma = L/\xi$ and $F_1 = (2 \xi/\xi_a) \exp(-L/\xi_a) \mbox{tr}\, t^{\dagger} t$ and take the limit $N \to \infty$. The resulting equation reads
\begin{equation}
  \partial_{\sigma} \langle F_1 \rangle = \delta_{\beta,1} (\langle F_1' \rangle - \case{3}{4} \langle F_1 \rangle), \label{eq:F1}
\end{equation}
where $F_1' = (2 \xi/\xi_a) e^{L/\xi_a} \mbox{tr}\, t^{\dagger} t r^{\dagger} r $. The evolution equation for $\langle F_1' \rangle$ in the thick-wire limit takes a particularly simple form,
\begin{equation}
  (\xi_a/\xi) \partial_{\sigma} \langle F_1' \rangle = -2 \langle F_1' \rangle + (1/2) \langle F_1 \rangle. \label{eq:F1prime}
\end{equation}
Here we used that $\langle \mbox{tr}\, (1 - r^{\dagger} r)^2 \rangle = \xi/2 \xi_a$ for $L \gg \xi_a$. The left-hand side of Eq.\ (\ref{eq:F1prime}) vanishes, so that $\langle F_1' \rangle = \langle F_1 \rangle/4$. Substitution into Eq.\ (\ref{eq:F1}) yields
\begin{equation}
  \partial_{\sigma} \langle F_1 \rangle = - \case{1}{2} \delta_{\beta,1}  \langle F_1 \rangle, \label{eq:F1total}
\end{equation}
We use Eq.\ (\ref{eq:tavgN}) for the initial condition, $\langle F_1 \rangle = 1$ for $\sigma \to 0$. (Notice that this is not the ballistic initial condition. Because of the order of limits taken, the initial condition $\sigma \to 0$ has to be evaluated within the strong absorption regime.) Hence
\begin{equation}
  \langle F_1 \rangle = \exp \left( - {L \delta_{\beta,1} / 2 \xi } \right).
\end{equation}
For $\xi_a \ll L \ll \xi$ this agrees with the weak localization correction to $\langle T \rangle$ in the diffusive regime for strong absorption, cf.\ Eq.\ (\ref{eq:tweak}).

Let us now consider the $L$-dependence of a general average $\langle \prod_{j=1}^{m} F_{n_{j}} \rangle$ in the thick-wire limit. Following the derivation of Eq.\ (\ref{eq:F1total}), we find the general evolution equation
\bleq
\begin{mathletters}
\begin{eqnarray} 
  \partial_{\sigma} \left \langle \prod_{j=1}^{m} F_{n_{j}} \right \rangle &=&
  \sum_{j=1}^{m} \sum_{i=1}^{n_{j}-1} {n_{j} \over 4} \left \langle F_{i} F_{n_{j}-i}  \prod_{{k=1 \atop k\neq j}}^{m} F_{n_{k}} \right\rangle 
  + \sum_{i < j}^{m} {n_{i} n_{j} \over 2} \left\langle F_{n_{i}+n_{j}} \prod_{{k=1 \atop k\neq i,j}}^{m} F_{n_{k}} \right\rangle 
  \nonumber \\ && \mbox{}
  - { \delta_{\beta,1}}
  \sum_{j=1}^{m} {n_{j} \over 2} \left \langle \prod_{{k=1}}^{m} F_{n_{k}} \right\rangle, \label{eq:evolTkkT} \\
  \left \langle \prod_{j=1}^{m} F_{n_{j}} \right\rangle_{\sigma \to 0} &=&
  \prod_{j=1}^{m} \delta_{n_{j},1}.
\end{eqnarray}
\end{mathletters}%
\eleq\noindent

The set of linear differential equations (\ref{eq:evolTkkT}) is readily solved. Its solution reads for $\sum_j {n_{j}} \le 3$
\begin{mathletters} \label{eq:evolTkkTsol}
\begin{eqnarray}
  \langle F_{1} \rangle &=& e^{- \delta_{\beta,1} \sigma/2}, \\
  \langle F_{1}^2 \rangle &=& \cosh \left({\sigma / 2} \right) e^{- \delta_{\beta,1} \sigma}, \\
  \langle F_{2} \rangle &=& \sinh \left({\sigma /2} \right) e^{- \delta_{\beta,1} \sigma}, \\
  \langle F_{1}^3 \rangle &=& \case{1}{3} \left[ \cosh \left({3\sigma /2} \right) + 2 \right]  e^{- 3\delta_{\beta,1} \sigma/2}, \\
  \langle F_{1} F_{2} \rangle &=& \case{1}{3} \sinh \left({3\sigma /2} \right) e^{- 3\delta_{\beta,1} \sigma/2}\\
  \langle F_{3} \rangle &=& \case{1}{3} \left[ \cosh \left({3\sigma /2} \right) - 1 \right] e^{- 3\delta_{\beta,1} \sigma/2}.
\end{eqnarray}
\end{mathletters}%
Deep in the localized regime $\sigma \gg 1$, the solution of Eq.\ (\ref{eq:evolTkkT}) has the asymptotic form
\begin{equation} \label{eq:asymploc}
  \log \langle F_1^{n} \rangle = \case{1}{4} \sigma \left[ n(n-1) - 2 n \delta_{\beta,1} \right] - \log n!.
\end{equation}

In the diffusive regime $L \ll \xi$, the fluctuations of the transmittance $T$ are much smaller than the average. The variance of the transmittance $T$ in the strong absorption regime can easily obtained from Eq.\ (\ref{eq:evolTkkTsol}) and agrees with the result of Sec.\ \ref{sec:3}. We also note that the third cumulant $\langle T^3 \rangle_c$ behaves as
\begin{mathletters}
\begin{eqnarray}
  \langle T^3 \rangle_c/\langle T \rangle^3 &=& \case{1}{16} (L/\xi)^4 + {\cal O}(L/\xi)^6,
\end{eqnarray}
\end{mathletters}%
which is qualitatively different from the case of a non-absorbing waveguide, where $\langle T^3 \rangle_c/\langle T \rangle^3 \sim (L/\xi)^5$ for $\beta=1$ and $ \sim (L/\xi)^6$ for $\beta=2$.\cite{Macedo,RossumLerner} 

For $L \gg \xi$, the fluctuations of $T$ are much larger than the average. Therefore, the average and variance of the transmittance are no longer sufficient to characterize the distribution. Instead, the transmission distribution is log-normal, like for a waveguide without absorption. This can be understood from the fact that for a log-normal distribution of $T$, the $n$th moment $\langle T^n \rangle$ reads 
\begin{equation} \label{eq:lognormal}
  \log \langle T^{n} \rangle = n \langle \log T \rangle + (n^2/2) \mbox{var}\, \log T.
\end{equation}
Using the asymptotic result (\ref{eq:asymploc}), together with Eq.\ (\ref{eq:Tndef}), we see that Eq.\ (\ref{eq:lognormal}) is satisfied for all $n$ with
\begin{mathletters}
\begin{eqnarray}
  \langle \log T \rangle &=& - L/\xi_a
   - {(L/4 \xi) \left( 1 + 2\, \delta_{\beta,1} \right)}, \\
  \mbox{var}\, \log T &=& L/2 \xi.
\end{eqnarray}
\end{mathletters}%
[We neglected the second $L$-independent term on the r.h.s.\ of Eq.\ (\ref{eq:asymploc}).\cite{footnote}] For a waveguide without absorption, on the contrary, the average and variance of $\log T$ are approximately equal and depend on the localization length $\xi$ only: $\langle \log T \rangle = - 2L/\beta \xi$ and $\mbox{var}\, \log T = 4 L/\beta \xi$. The insensitivity to $\beta$ of the transmittance fluctuations for a waveguide with absorption was already discussed in Sec.\ \ref{sec:3}.

\section{Transmittances $T_{a}$ and $T_{ab}$} \label{sec:5}

So far we have mainly considered the statistical distribution of the transmittance $T$, which is appropriate if the waveguide is illuminated through a diffusor. Using the results of the previous two sections, it is only a small step to find the distributions of the transmittances $T_{a}$ and $T_{ab}$, which describe a situation in which the waveguide is illuminated through a single channel only (e.g.\ by a plane wave). In fact, it is sufficient to consider the transmittance $T_{a}$, since the distribution of $T_{ab}$ follows directly from that of $T_{a}$,\cite{KoganKavehBaumgartnerBerkovits}
\begin{equation}
  P(T_{ab}) = N \int_0^{\infty} {dT_{a} \over T_{a}} P(T_{a}) \exp \left( - {N T_{ab} \over T_{a}} \right). \label{eq:TabTa}
\end{equation}

The statistical distribution of the transmittances $T_a$ and $T_{ab}$ for non-absorbing random waveguides has been calculated in Refs.\ \onlinecite{NieuwenhuizenRossum,KoganKaveh,vanLangen}. For $L \ll \xi$, the distribution of $T_{a}$ is Gaussian with mean $\langle T_{a} \rangle = N^{-1} \langle T \rangle = \ell/L$, variance $\mbox{var}\, T_{a} = 2 \ell/3NL$, and non-Gaussian tails. The non-Gaussian features of the distribution become more pronounced as the length of the waveguide approaches the localization length $\xi = N \ell$. In the localized regime $L \gg \xi$, $T_{a}$ has a log-normal distribution, with the same mean and variance as the transmittance $T$.\cite{vanLangen}

Let us now discuss the effect of absorption on the distribution of $T_{a}$. (The variance of $T_a$ in the presence of absorption has also been considered in Refs.\ \onlinecite{Stephen} and \onlinecite{PniniShapiro}.) In terms of the unitary matrix $u$ that diagonalizes $t t^{\dagger}$ [cf.\ Eq.\ (\ref{eq:polar})] and the transmission eigenvalues $\tau_{\mu}$, the transmittance $T_{a}$ reads
\begin{equation}
  T_{a} = \sum_{\mu} |u_{a \mu}|^2 \tau_{\mu}.
\end{equation}
The matrix $u$ is uniformly distributed in the unitary group. For large $N$, we may consider the matrix elements $u_{a \mu}$ as independently distributed Gaussian random numbers with zero mean and variance $\langle |u_{a \mu}|^2 \rangle = 1/N$. In this limit, the Laplace transform $F(z)$ of $P(T_a)$ becomes\cite{KoganKaveh,vanLangen}
\begin{eqnarray}
  F(z) &=& \int_0^{\infty} dT_{a} e^{-N z T_a} P(T_a) \nonumber \\ &=&
           \left \langle \prod_{\mu} (1 + z \tau_{\mu})^{-1} \right\rangle.
\end{eqnarray}

In the strong absorption regime $L \gg \xi_a$, we can substitute the results of Sec.\ \ref{sec:4} for the averages of moments of the transmission eigenvalues. The result is
\begin{eqnarray}
  F(z) &=& \sum_{n=0}^{\infty} z^n \exp \left[ n(n-1) {L \over 4 \xi} - n \delta_{\beta,1} {L \over 2 \xi} - n {L \over \xi_a} \right].
\end{eqnarray}
Transforming back, we find that $T_a$ is log-normally distributed with mean and variance
\begin{mathletters} \label{eq:TaLogNormal}
\begin{eqnarray}
  \langle \log T_a \rangle &=& - {L \over \xi_a} - (1 + 2 \delta_{\beta,1}) {L \over 4 \xi} - \log N,\\
  \mbox{var}\, \log T_a &=& {L \over 2 \xi}.
\end{eqnarray}
\end{mathletters}
The origin of the remarkable simple log-normal distribution of $T_a$ in the strong absorption regime is understood from a simple argument.\cite{Shnerb} In the strong absorption regime, diffusion takes place only inside a volume $\xi_a \ll L$, and motion is quasi-ballistic on larger length scales. Inside a segment of $\xi_a$, absorption plays no role, and the distribution of the transmittance $T_a$ is Gaussian, with mean $\sim \ell/\xi_a$ and variance $\sim \ell^2/\xi \xi_a$, as in the case of a non-absorbing random waveguide. Since the dynamics on larger length scales is quasi-ballistic, the transmittance $T_a$ of a waveguide of length $L$ is the product of the individual transmittances of uncorrelated segments of size $\xi_a$. From the central limit theorem we then immediately derive that for $L \gg \xi_a$, the transmittance $T_a$ is log-normally distributed with variance $\mbox{var}\, \log T_a \sim L/\xi$.
\begin{figure}

\epsfxsize=0.89\hsize
\hspace{0.00\hsize}
\epsffile{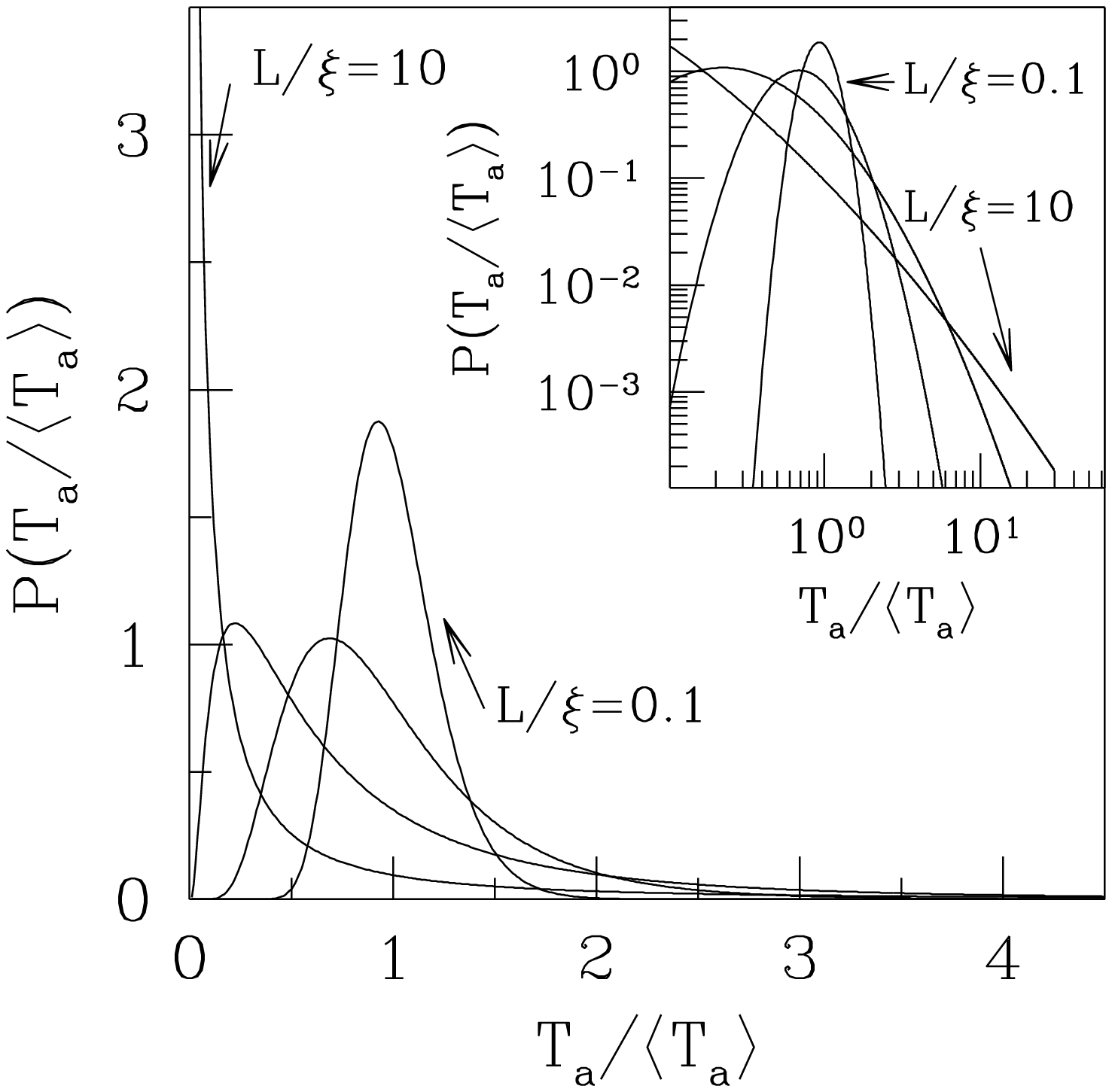}
\refstepcounter{figure} 
\label{fig:Ta}

{\small FIG.\ \ref{fig:Ta}. Distribution of the transmittance $T_a$ for $L/\xi = 0.1$, $0.5$, $2$, and $10$ in a strongly absorbing random waveguide. Inset: The same, but on a logarithmic scale.\par}
\end{figure}

The distribution of $T_a$ in the strong absorption regime is shown in Fig.\ \ref{fig:Ta} for several values of $L/\xi$. In the localized regime $L \gg \xi$, the distribution of $N T_a$ is the same as that of the transmittance (or conductance) $T$. However, the log-normal distribution (\ref{eq:TaLogNormal}) of the transmittance $T_a$ is valid for all $L \gg \xi_a$, i.e.\ before the onset of localization. 

In the diffusive regime $\ell \ll L \ll \xi$, the distribution of $T_a$ is sharply peaked around the average $\langle T_a \rangle = N^{-1} \langle T \rangle$. Fluctuations around the average are described by the cumulants
\begin{mathletters} \label{eq:TaCumul}
\begin{eqnarray}
  \mbox{var}\, T_a &=& {\xi_a \over 4 \xi} \langle T_a \rangle^2 \left(
  2 s + \coth s  - {s \over \sinh^2 s} \right), \\
  \langle T_a^3 \rangle_c &=&
  {3 \xi_a^2 \over 32 \xi^2} \langle T_a \rangle^3 \left( 8 s^2 + 9 + {4 s^2 \over \sinh^4 s}  \right. \nonumber \\ && \hphantom{{3 \xi_a^2 \over 32 \xi^2} \langle T_a \rangle^3} \mbox{} \left. - {6 s^2 + 5 s \coth s - 1 \over \sinh^2 s}\right),
\end{eqnarray}
\end{mathletters}%
where $s = L/\xi_a$. Eq.\ (\ref{eq:TaCumul}) is valid for both weak and strong absorption. In the weak absorption regime $L \ll \xi_a$, Eq.\ (\ref{eq:TaCumul}) reduces to the known results for a non-absorbing waveguide\cite{NieuwenhuizenRossum,KoganKaveh}
\begin{eqnarray} \label{eq:momNonAbs}
  \mbox{var}\,T_{a} &=& {2 L \over 3 \xi}\langle T_{a} \rangle^2, \ \
  \langle T_{a}^3 \rangle_{c} = {16 L^2 \over 15 \xi^2}\langle T_{a} \rangle^3.
\end{eqnarray}
whereas for strong absorption we find
\begin{eqnarray} \label{eq:momAbs}
  \mbox{var}\, T_{a} &=& {L \over 2 \xi} \langle T_{a} \rangle^2,\ \
  \langle T_{a}^3 \rangle_{c} = {3 L^2 \over  4 \xi^2} \langle T_{a} \rangle^3.
\end{eqnarray}
The strong absorption limit of $\mbox{var}\, T_a$ was previously obtained in Refs.\ \onlinecite{PniniShapiro,KoganKaveh92,GenackShapiro}.

\begin{figure}

\epsfxsize=0.89\hsize
\hspace{0.00\hsize}
\epsffile{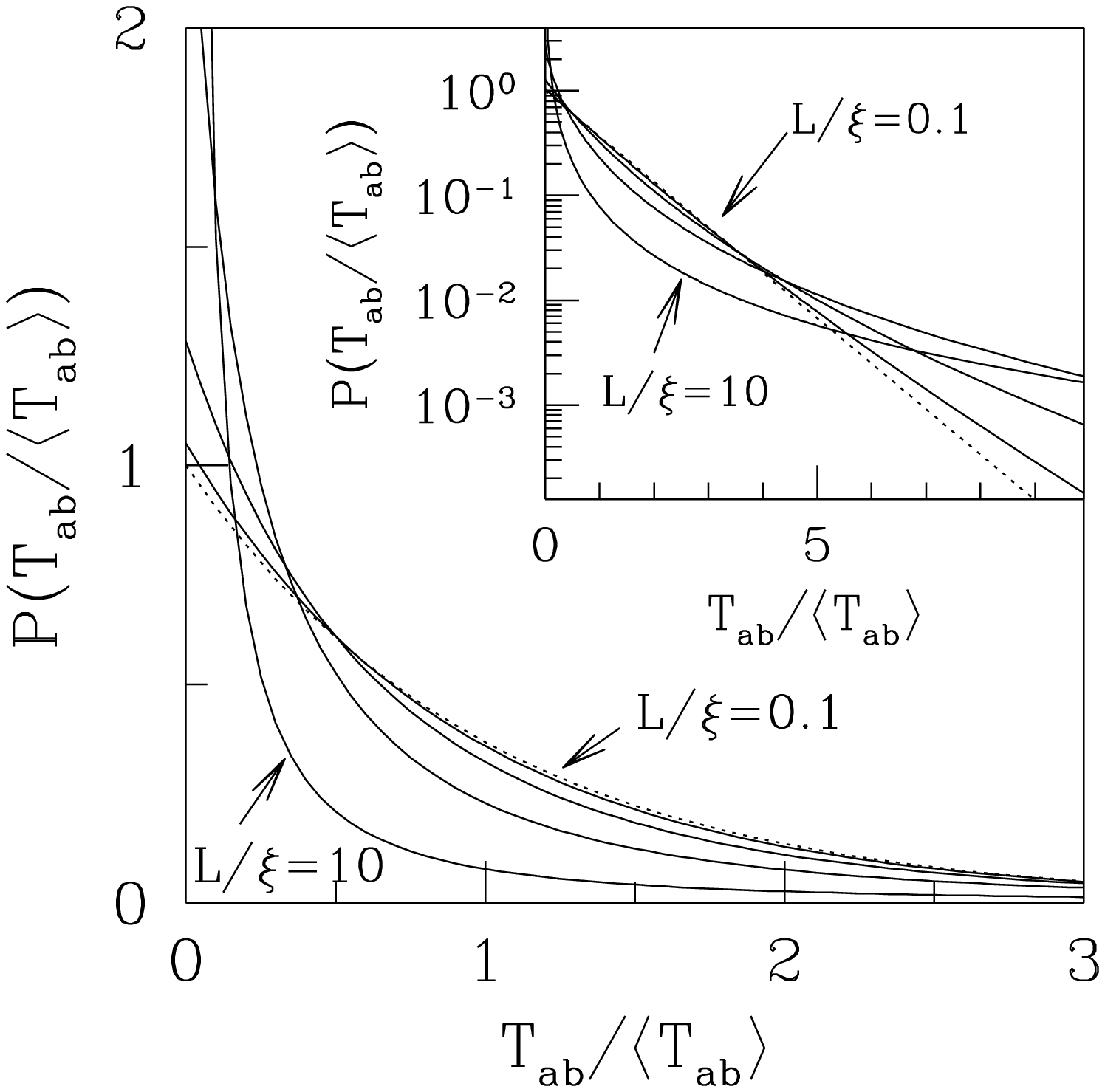}
\refstepcounter{figure} 
\label{fig:Tab}

{\small FIG.\ \ref{fig:Tab}. Distribution of the transmittance $T_{ab}$ for $L/\xi = 0.1$, $0.5$, $2$, and $10$, in a strongly absorbing random waveguide. The limiting negative exponential distribution for $L \ll \xi$ is shown dotted.\par}
\end{figure}

The distribution of the transmittance $T_{ab}$ in the strong absorption regime is easily obtained using Eq.\ (\ref{eq:TabTa}). It is shown in Fig.\ \ref{fig:Tab} for several values of $L/\xi$. For $L \ll \xi$, the distribution of $T_{ab}$ is negative exponential (Rayleigh). In the localized regime $L \gg \xi$, the distribution is log-normal.

\section{Experiments} \label{sec:5b}

In this section, we discuss the application of our work to the recent microwave experiments by Stoytchev and Genack.\cite{Stoytchev} In these experiments, the distribution of the transmittance $T_a$ was measured in a cylindrical copper tube with randomly placed polystyrene scatterers. The presence of many scatterers in a narrow waveguide results in a relatively small localization length ($\xi \sim 5$m, while the length $L$ of the waveguide varies between $50$cm and $2$m), but at the cost of strong absorption ($\xi_a \approx 30$cm). Lacking a theory for absorbing random waveguides, the authors of Ref.\ \onlinecite{Stoytchev} considered the distribution of the transmittance divided by its mean, ${\cal T}_a = T_a/\langle T_a \rangle$, and compared it to the theoretical predictions for non-absorbing waveguides.\cite{NieuwenhuizenRossum,KoganKaveh,vanLangen} Because of the absorption, the average transmittance $\langle T_a \rangle$ can not be used to determine the ratio $L/\xi$. Instead, in Ref.\ \onlinecite{Stoytchev}, $L/\xi$ was computed from the formula $\mbox{var}\, {\cal T}_a = (2/3) (L/\xi)$, which is valid for weakly absorbing waveguides only [cf.\ Eq.\ (\ref{eq:momNonAbs})]. 

Surprisingly, the measured probability distribution of ${\cal T}_a$ was found to follow the theoretical predictions for non-absorbing waveguides in the diffusive regime quite accurately, provided one uses the ratio $L/\xi$ obtained from $\mbox{var}\, {\cal T}_a$, as explained above. This is surprising, because the waveguides of Ref.\ \onlinecite{Stoytchev} are strongly absorbing, $L/\xi_a$ ranging from $2$ to $6$, and because the experiment is close to the localization threshold ($\xi/L$ up to $3$). The agreement is not so good if $L/\xi$ is directly extracted from the experimental parameters. In particular, the measured variance of ${\cal T}_a$ was found to depend sublinearly on $L$, while the theory for a non-absorbing waveguide predicts a linear $L$-dependence (or a superlinear $L$-dependence if localization effects are taken into account). 

\begin{figure}

\epsfxsize=0.89\hsize
\hspace{0.00\hsize}
\epsffile{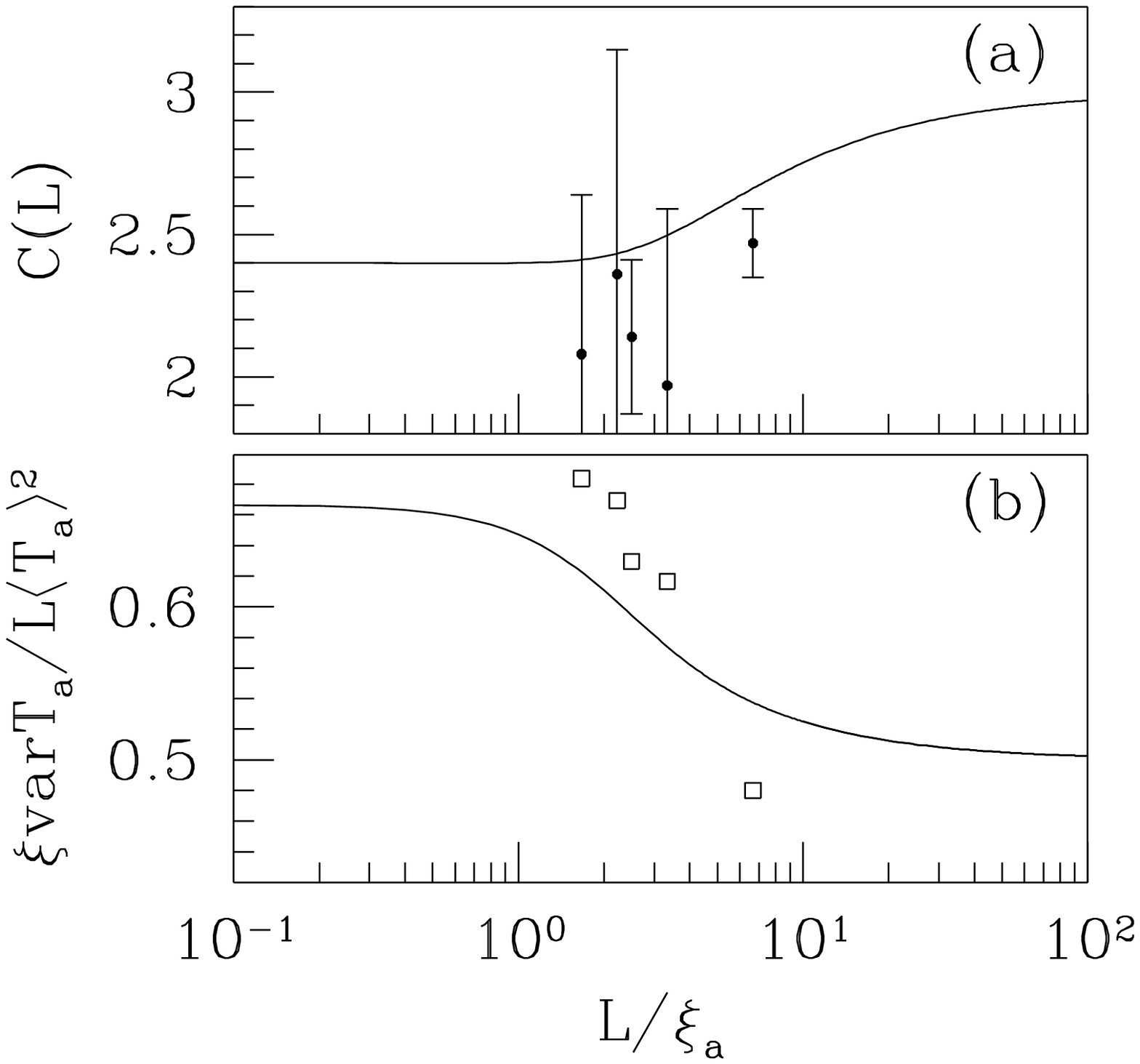}
\refstepcounter{figure} 
\label{fig:VarTa}

{\small FIG.\ \ref{fig:VarTa}. (a) The ratio $C(L) = \langle {\cal T}_{a}^3 \rangle_c /(\mbox{var}\, {\cal T}_a)^2$ versus $L/\xi_a$ in the crossover between weak and strong absorption. (b) The ratio $(\xi/L) \mbox{var}\, {\cal T}_a$ versus $L/\xi_a$. The errorbars (a) and squares (b) indicate the results of the experiment of Ref.\ \protect\onlinecite{Stoytchev}.\par}
\end{figure}

How do these observations compare to our results for the transmittance distribution of random waveguides with absorption? Following Ref.\ \onlinecite{Stoytchev}, we use the ratio $C(L) = \langle {\cal T}_{a}^3 \rangle_c/(\mbox{var}\, {\cal T}_a)^2$ to characterize the distribution.  The theoretical predictions for $C(L)$ and $(\xi/L) \mbox{var}\, {\cal T}_a$ as a function of $L/\xi_a$ in the diffusive regime are shown in Fig.\ \ref{fig:VarTa} [see also Eq.\ (\ref{eq:TaCumul})]. We find that 
\begin{itemize}
\item as a result of absorption, $\mbox{var}\, {\cal T}_a$ crosses over from $\case{2}{3} L/\xi $ to $\case{1}{2} L/\xi$ for $1 \lesssim L/\xi_a \lesssim 10$,
\item the ratio $C(L)$ crosses over from $C(L) = 12/5$ for weak absorption to $3$ for strong absorption, the crossover occurring for $3 \lesssim L/\xi_a \lesssim 30$.
\end{itemize}
The length $L$ of the random waveguides in the experiment lies between $2\, \xi_a$ and $6\, \xi_a$. Hence, the waveguides are long enough to fully observe the sublinear behavior of $\mbox{var}\, {\cal T}_a$, but too short to see a significant enhancement of $C(L)$ above the weak absorption limit $C(L) = 12/5$. (Note that the experiment indeed shows a slight enhancement of $C(L)$ for the longest waveguide.\cite{Stoytchev})
We expect that higher cumulants of ${\cal T}_a$, if properly normalized, show the same ``postponed'' crossover behavior as $\langle {\cal T}_a^3 \rangle_c$. This could explain why the entire transmittance distribution of Ref.\ \onlinecite{Stoytchev} agrees with weak-absorption theory rather than with strong-absorption theory. 

While the above considerations offer a qualitative explanation for the most striking experimental observations, the theory presented here fails to account for the measurement of $\mbox{var}\, {\cal T}_a$ versus $L$ quantitatively. This is clearly illustrated by comparison of the experimental results of Ref.\ \onlinecite{Stoytchev} and the theory in Fig.\ \ref{fig:VarTa}b. Notice that the fact that the theoretical curves in the figure were derived for $L \ll \xi$, while the ratio $\xi/L$ is not necessarily large in the experiment ($\xi/L$ down to $3$), cannot explain the difference, as localization effects increase fluctuations rather than decrease them. For the longest waveguide ($L/\xi_a \approx 6$), Stoytchev and Genack find $\mbox{var}\, {\cal T}_a \approx 0.43 L/\xi$, whereas we find $\mbox{var}\, {\cal T}_a \ge \case{1}{2} L/\xi$ for all $L$. The reason for this discrepancy is not known. A quantitative comparison with the experiment for $C(L)$ is difficult because of the uncertainty of the experimental data (see Fig.\ \ref{fig:VarTa}a).

According to Eq.\ (\ref{eq:TaCumul}) or Fig.\ \ref{fig:VarTa}a, absorption causes the distribution of the transmittance to deviate significantly from that of a non-absorbing waveguide only if $L/\xi_a \gtrsim 10$. It should not be difficult to verify this experimentally, e.g.\ by a decrease of $\xi_a$ due to the addition of strongly absorbing scatterers to the waveguide. We also find that the size of the fluctuations depends linearly on $L/\xi$, both in the weak and strong absorption regimes, but with different slopes ($2/3$ and $1/2$, respectively). While the experiment of Ref.\ \onlinecite{Stoytchev} confirms the initial $L$-dependence with slope $2/3$ as well as the deviation for $L \sim \xi_a$, more experimental input is required to verify the linear dependence with slope $1/2$ in the strongly absorbing regime. 

\section{Conclusion} \label{sec:6}

We have computed the statistical properties of the transmittance of a multichannel random waveguide with absorption. We have considered both the diffusive regime, where the transmittance distribution is Gaussian and the localized regime, where the transmittance distribution is log-normal. Our main findings are summarized in Table \ref{table:1}. 
\begin{center}
\begin{tabular}{c|c|c}
  & no $/$ weak & strong \\
  & absorption & absorption \\ 
  \hline\hline
  diffusive &
  $\ell \ll L \ll \xi_a$ & $\xi_a \ll L \ll \xi$ \\ \hline
  $P(T)$ & 
  Gaussian &
  Gaussian \\
  $\langle T \rangle$ & 
  $\xi/L$ & 
  $(2\xi/\xi_a) e^{-L/\xi_a}$ \\
  $\delta T$ &
  $-\case{1}{3} (L/\xi) \langle T \rangle$ &
  $-\case{1}{2} (L/\xi) \langle T \rangle$ \\
  $\mbox{var}\, T$ & 
  $~{2 \over 15 \beta} (L/\xi)^2 \langle T \rangle^2$~ &
  ${1 \over 8} (L/\xi)^2 \langle T \rangle^2$ \\
  $\rho(\tau)$ &
  bimodal &
  $~~ \tau_{\rm max} = e L \langle T \rangle/2 \xi~~ $ \\ \hline \hline
  localized & $L \gg \xi$ & $L \gg \xi$ \\ \hline
  $P(T)$ & log-normal & log-normal \\
  $\langle \log T \rangle$ &
  $-2L/\beta \xi$ &
  $-L/\xi_a + {\cal O}(L/\xi)$ \\
  $\mbox{var} \log T$ &
  $4L/\beta \xi$ &
  $L/2\xi$ 
\end{tabular}
\refstepcounter{table}
\label{table:1}
\end{center}
{\small TABLE \ref{table:1}. Statistical properties of the transmittance $T$ for weak and strong absorption. The relevant length scales are the the localization length $\xi = N \ell$ and the exponential decay length $\xi_a = [\ell \ell_a / 2]^{1/2}$, where $N$ is the number of channels, $\ell$ the elastic mean free path, and $\ell_a$ the ballistic absorption length. \par\bigskip}

To present, no optical or microwave experiments have been able to address the localized regime $L \gtrsim \xi$, where the fluctuations of the transmittance are larger than the average. The mere presence of strong absorption modifies the localization transition, but does not suppress the fluctuations. As stronger scatterers tend to be stronger absorbers, it might be easier from the experimental point of view to construct a random waveguide with $L \gtrsim \xi$ in the strong absorption regime than a comparable random waveguide without absorption.

Our work assumes that absorption sets in before localization, i.e.\ that $\xi_a \ll \xi$. The assumption was necessary, because we use $\xi_a/\xi$ as a small parameter. It is appropriate for the experiment of Stoytchev and Genack, in which the ratio $\xi_a/\xi$ is less than $0.1$. A theory of the localization transition in random waveguides that is nonperturbative in $\xi_a/\xi$, which would require a non-linear $\sigma$-model formulation,\cite{Efetov1983} remains to be developed.

To conclude, we would like to remark that the problem of transmission through random media in the presence of absorption is also important for the problem of ``directed localization''.\cite{HatanoNelson,Efetov,BSB,NelsonShnerb} Directed localization refers to the localization transition in a disordered ring with an imaginary vector potential, first introduced by Hatano and Nelson.\cite{HatanoNelson} This model is relevant for e.g.\ the pinning of vortices to columnar defects in superconductors, or for problems in population biology.\cite{NelsonShnerb} 
In a purely one-dimensional geometry ($N=1$), the support of the spectrum in the complex plane is entirely determined by the transmittance of the disordered ring at complex energies, i.e.\ with absorption (or gain).\cite{BSB} Such a relation may also exist for the multichannel waveguides considered in this paper.

\acknowledgments

We thank C.\ W.\ J.\ Beenakker, T.\ Sh.\ Misirpashaev, and N.\ M.\ Shnerb for stimulating discussions. We thank A.\ Z.\ Genack and M.\ Stoytchev for correspondence and communication of the experimental data. This work was supported by the ``Stich\-ting voor Fun\-da\-men\-teel On\-der\-zoek der Ma\-te\-rie'' (FOM) and by the ``Ne\-der\-land\-se or\-ga\-ni\-sa\-tie voor We\-ten\-schap\-pe\-lijk On\-der\-zoek'' (NWO) and by the NSF under grants no.\ DMR 94-16910, DMR 96-30064, and DMR 94-17047.

\ecols


\begin{references}
\bibitem{footnote} If we take the $L$-independent terms of 
         Eqs.\ (\ref{eq:Tndef}) and (\ref{eq:asymploc}) into account, we find 
$$
  \langle \log T \rangle =  - {L \over \xi_a}
   - {L \over 2 \xi} \left( 1 + 2\, \delta_{\beta,1} \right) +  \log {2 \xi \over \xi_1} + c,
$$
where $c \approx 0.57$.
\bibitem{Shnerb} I learned this argument from N.\ M.\ Shnerb. 
         It is valid for the transmittances $T_a$, because these 
         satisfy a multiplicative law in absorbing systems for 
         $L \gtrsim \xi_a$. It does not apply to the transmittances 
         $T$ and $T_{ab}$, which have more complicated distributions 
         in the diffusive regime $L \lesssim \xi$.
\end{references}
\end{document}